# Nanomanufacturing of titania interfaces with controlled structural and functional properties by supersonic cluster beam deposition

Alessandro Podestà[1], Francesca Borghi, Marco Indrieri, Simone Bovio, Claudio Piazzoni, Paolo Milani[1]

Centro Interdisciplinare Materiali e Interfacce Nanostrutturati (C.I.Ma.I.Na.),

Dipartimento di Fisica, Università degli Studi di Milano, via Celoria 16, 20133 Milano, Italy.

[1] Corresponding authors. E-mail: alessandro.podesta@mi.infn.it; pmilani@mi.infn.it

**Keywords.** Supersonic Cluster Beam Deposition; Pulsed Micro-plasma Cluster Source; nanostructured titania; surface morphology; scaling laws; Atomic Force Microscopy.

**Abstract.** Great emphasis is placed on the development of integrated approaches for the synthesis and the characterization of ad hoc nanostructured platforms, to be used as templates with controlled morphology and chemical properties for the investigation of specific phenomena of great relevance for technological applications in interdisciplinary fields such as biotechnology, medicine and advanced materials. Here we discuss the crucial role and the advantages of thin film deposition strategies based on cluster-assembling from supersonic cluster beams. We select cluster-assembled nanostructured titania (ns-$TiO_2$) as a case study to demonstrate that accurate control over morphological parameters can be routinely achieved, and consequently over several relevant interfacial properties and phenomena, like surface charging in a liquid electrolyte, and proteins and nanoparticles adsorption.

**Introduction**

Titanium dioxide nanostructured films assembled by nanoparticles are widely used for photocatalysis[1–4] and solar energy conversion[5–8]. These applications require layers with very large accessible active or specific area, usually several hundred square meters per gram of material, in order to optimize the mass transfer rates for catalytic applications[9] and to increase the electron diffusion pathways in dye-sensitized solar cells[5,10].

Several techniques are currently applied to the large-scale production of $TiO_2$ nanoparticles and their subsequent assembly: sol–gel processing[7], hydrolysis[11], physical and chemical vapor deposition[12], flame pyrolysis[13]. Wet chemistry approaches are very effective for large-scale production, however they present substantial limitations about the building of layers and interfaces with well controlled porosity, morphology and lateral resolution[14].

Gas phase routes for the production of nanostructured titania porous layers and interfaces offer several advantages due to the possibility of a precise control on physico-chemical properties such as the phase, morphology and degree of agglomeration of nanoparticles prior to deposition[14–16]. In particular the degree of agglomeration of primary spherical nanoparticles forming fractal aggregates in the gas phase prior to deposition and the landing kinetic energy on the substrate are parameters affecting the porosity of the layer[16].

$TiO_2$ is also one of the most widely used biocompatible materials for dental and orthopedic prosthetics[17–20]: the role of the nano and microscale surface morphology in determining the efficacy of the implant has been widely recognized and a conspicuous literature exists about the effect of titania surface nanostructure on cell adhesion, spread, growth and differentiation[21–23]. Despite the remarkably large number of published reports, a coherent picture about the role of nanoscale structural features for the control of complex biological phenomena at the biotic/abiotic interface is not yet emerging. One possible reason of this situation is that a systematic evaluation of the surface morphological and topographical cues relevant for the interaction with biological entities requires the exploration of a multi-dimensional production parameter space and hence the use of high-throughput and reproducible synthetic methods over a wide range of dimensions (from the macro-to the nanoscale).

The fabrication of titania nanostructured surfaces is usually performed by subtractive top-down methods such as chemical etching, sand blasting or microfabrication approaches typical of silicon technologies[24]. All these approaches fail in capturing the morphological complexity typical of the extracellular matrix which is recognized as one of key parameters in affecting cell adhesion, proliferation and differentiation[25]. Anodization techniques using porous aluminum anodic masks have been shown to be effective in producing assemblies of titania nanopillars with nanoscale features that can be systematically varied; however this approach cannot be used to integrate nanostructured surfaces onto microfabricated platforms[26].

Recently a series of papers have been published demonstrating that cluster-assembling from the gas phase and, in particular, supersonic cluster beam deposition (SCBD) can produce nanostructured titania films with tailored and reproducible nanoscale surface morphologies for protein, enzyme and virus adsorption[27,28], cellular and bacterial adhesion, proliferation and differentiation[29,30].

Here we show that SCBD can be used as an efficient additive method for the prototyping of nanostructured interfaces with controlled morphological and functional properties. The results reported in this work refer to a very large number of samples produced in a time window of approximately 10 years with different SCBD apparatus in different academic and industrial facilities. We report the extensive and systematic characterization of ns-$TiO_2$ films deposited by SCBD in different conditions, highlighting the mechanisms of film growth determining the observed regularity and reproducibility of the deposition process. We discuss the implications of these mechanisms on the possibility of controlling by design the properties of the deposited thin films, presenting examples of applications where nanoscale morphology has a strong and direct impact on the structural and functional properties of the surface.

1. **Experimental**

*1.1 Deposition of ns-$TiO_2$ films by Supersonic Cluster Beam Deposition*

The fundamental tool for the synthesis of thin films with controlled nanoscale morphology by cluster-assembling is an SCBD apparatus equipped with a Pulsed Micro-plasma Cluster Source (PMCS)[15,31]. Details on the deposition technique are provided elsewhere[15,31]; here we summarize and highlight the most important aspects for the control of morphological properties and interfacial phenomena. The SCBD apparatus is shown schematically in Figure 1. The apparatus consists of three differentially pumped vacuum chambers. The first stage is an expansion chamber where the supersonic molecular beam is formed; it can be connected to a second chamber by an electroformed skimmer. The cluster deposition takes place in a third chamber connected to the rest of the apparatus through a gate valve. A PMCS is mounted outside the expansion chamber on the axis of the apparatus. A remotely controlled manipulator allows for rastering of the sample to guarantee a uniform deposition over a large area. The structure of a PMCS is shown in the inset of Figure 1: it schematically consists of a ceramic body (2) with a channel drilled to perpendicularly intersect a larger cylindrical cavity. The channel hosts a titanium target rod (3) acting as a cathode in order to produce the cluster precursors. A solenoid pulsed valve (1) faces one side of the cavity and a removable nozzle closes the other side of it. The valve, backed with a high inert gas pressure (20-50 bars), injects in the source cavity pulses with duration of few hundreds of microseconds at a repetition rate of 3-10 Hz. The nozzle is connected with a series of aerodynamic lenses (4) used to focus neutral nanoparticles on the beam axis.

The pulsed injection of the inert carrier gas in the cavity of the PMCS causes the formation of a supersonic jet directed against the target rod. Synchronous with the gas injection, a pulsed voltage (50-100μs of duration, 700-1000 V) is applied to the target cathode and the grounded anode (pulsed valve front) in order to ionize the gas and sputter the target. Due to the plasma confinement obtained by the aerodynamic effects of supersonic expansions, the sputtering process is very efficient and reproducible[32]. The species ablated from the target thermalize with the inert gas and condense to form clusters. The carrier gas-cluster mixture expands out of the nozzle forming a seeded supersonic expansion,[33] which impinges on the substrate holder in the deposition chamber.

Clusters are not monodispersed in size; instead, they possess a rather broad size distribution when they exit from the PMCS, which depends on the carrier gas and on the operational parameters of the source[31]. Aerodynamic focusing based on inertial effects of clusters, obtained by using special nozzle configurations, filters off the largest clusters and aggregates, and concentrate particles along the beam axis. The cluster beam profile is approximately gaussian, with larger particles concentrated at the beam center; nanoparticles diameter decreases, going from the beam center to the periphery[33]. By controlling the working parameters of the PMCS, the aerodynamic filters, and the portion of the beam intercepting the substrate, the nanoparticles distribution can be precisely tuned and reproduced. Consequently, once the SCBD parameters are set, the deposition time controls the surface morphology of the films, which evolves regularly according to simple and reproducible scaling laws.

Ns-TiO$_x$ (x≤2) clusters have been produced in the PMCS using Ar or He as carrier gas, then deposited on suitable substrates, typically borosilicate glass coverslips (diameter 15 mm, thickness 0.13-0.17 mm) or fragments of polished Si wafers, intercepting the beam in the deposition chamber. Ti clusters partially oxidize in the cluster source and in the deposition chamber; oxidation further proceeds upon exposure of the film to air, and completes after annealing at 250°C[34].

Ns-TiO$_2$ samples are characterized by thickness in the range 5-250 nm, rms roughness $w$ ranging from 5 to 30 nm and specific area $r$ from 1.2 to 1.9 (morphological parameters are introduced in the next section).

*1.2 Characterization of surface morphology of ns-TiO$_2$ films by Atomic Force Microscopy*

The investigation of the morphology of the substrates has been carried out in air using both Multimode and Bioscope Catalyst atomic force microscopes (AFM) from Bruker equipped with Nanoscope IV and Nanoscope V controllers, respectively. The AFM were operated in Tapping Mode using single-crystal silicon tips with nominal radius of curvature in the range 5–10 nm and cantilever resonance frequency in the range of 200–300 kHz. Scan area was typically 2μm x 1μm with scan rates of 1.5–2 Hz. Sampling resolution was 2048x512 points. Due to its photo-catalytic activity, ns-TiO$_2$ gets quickly contaminated upon exposure to air by organic and other contaminants, as well as water[34]; therefore, prior to imaging, ns-TiO$_2$ samples were typically cured for 2 hours at 250°C in air in order to recover a clean, hydroxylated surface[35]. The thermal treatments at temperatures below 400°C cause negligible relative changes of the values of morphological parameters, of the order of 1%.

Several AFM images were typically acquired on each sample. Images were flattened by line-by-line subtraction of first and second order polynomials in order to remove artifacts due to sample tilt and scanner bow. Figure 2 shows a schematic representation of a topographic one-dimensional profile h(x), with all the relevant quantities and their geometrical meaning highlighted. In particular, from flattened AFM images the root-mean-square surface roughness $w$ (the vertical width of the surface) is calculated as the standard deviation of surface heights; specific area $r$ is calculated as the ratio of surface area to the projected area (in Figure 2, where the one-dimensional case is represented, this corresponds to the ratio of profile lengths). The specific area calculated from AFM images is always underestimated because of the inability of the AFM tip to detect overhangs and because of its finite size, causing tip-sample convolution effects[36]. Rms roughness for relatively rough surfaces like those considered in this work is typically not significantly affected by convolution effects. A surface pore is highlighted (shaded in red) in Figure 2. The correlation length ξ, i.e. the horizontal width of the surface, is also indicated; ξ represents a statistical measure of the halved average lateral dimension of the largest morphological surface features. The film thickness was calculated by scanning a region of the film where a sharp step was produced by masking the substrate during cluster deposition.

Correlation functions C(l) and G(l), defined later in the text, were evaluated by first averaging over all pairs of points (taken along the profiles) in a topographic map separated by a distance L, then averaging curves from different topographic maps of the same sample.

The cluster size distributions were characterized by depositing clusters on smooth flat substrates (glass coverslips or freshly cleaved mica surfaces) at very low coverage, and imaging them by AFM. In high-resolution AFM topographies, a few hundred isolated particles can be detected, labeled, and morphological parameters such as height, volume, projected area, can be measured according to custom data-analysis protocols for particle sizing. In particular, assuming particles possess an approximately spherical shape, the particle's height can be considered as an accurate representation of its diameter; the measurement of heights indeed is not affected by tip/sample convolution issues, as long as particles are well separated on the surface[36].

## 2. Results and Discussion

*2.1 Analysis of cluster size distribution by AFM*

Figures 3(a-d) show representative AFM images of isolate ns-TiO$_2$ clusters deposited using He and Ar on glass coverslips, intercepting the center and the periphery of the beam, respectively. The corresponding distributions of particle diameters characterized by AFM are shown in Figures 3(e-

h). Distributions are multimodal, each mode being approximately lognormal, therefore appearing as a Gaussian curve in semilog scale. The size distributions of $TiO_2$ nanoparticles produced using Ar and He are mainly bimodal: the first relevant difference between Ar and He is the median size of particles belonging to the major mode: the median diameter is 4.5 nm for Ar, and 2.1 nm for He. Selecting the carrier gas therefore allows shifting by a significant amount of the median particles' diameter. Inertial effects of clusters in the supersonic beam determine the concentration of larger particles along the beam axis, as proved by the depletion of the large-diameter mode in the case of Ar; in the case of He, depletion is less important, probably because particles in the major mode are already relatively small. The central portion of the cluster beam, the more intense, provides the greatest contribution to thin film growth; therefore Figures 3(e,g) represent a quantitative characterization of the size and relative abundance of building blocks used to produce ns-$TiO_2$ films.

*2.2 General description of surface nanoscale morphology*

Figures 4(a,b) show representative topographic maps of ns-$TiO_2$ films with similar thickness (about 100 nm) deposited using He and Ar as carrier gas, with rms roughness w of 8 and 21 nm, respectively. Despite thickness is similar, a marked difference in surface corrugation is observed, as well as in the average size of surface grains, which represent either the primeval clusters in the beam (yet dilated by tip convolution effects), or their aggregation/coalescence at the surface upon deposition. This difference can be explained in terms of the marked differences in the cluster size distributions obtained using He and Ar (Figure 3). Figures 4(a,b) also clearly show the granular, nanoporous nature cluster-assembled materials obtained by low-energy deposition, and the remarkable gain in specific area due to the use of small nanometer-sized building blocks. To further highlight these morphological properties, three-dimensional views of the surface of ns-$TiO_2$ films deposited using Ar, with different roughness of 10 and 22 nm (thickness of about 20 and 100 nm)

are shown in Figures 4(c,d). Clearly visible is the pattern of nanometer-sized grains and pores. Overall, the film is characterized by high specific area and porosity at the nano and sub-nanometer scale, extending in the bulk of the film. Surface pores and surface specific area, as well as rms roughness, depend on film thickness[27,35], and increase with it. The surface sections superimposed to the height maps in Figures 4(c,d) show nanometric pores of diverse depths and widths. As the film thickness increases, not only the surface becomes rougher, but also the average lateral dimension of the largest morphological features (the correlation length $\xi$) increases. An evolution of the aspect ratio of surface pores is therefore to be expected. We have recently shown that the increase of surface roughness w determines the increase of the number of pore having higher aspect ratio; moreover, as w increases, also the average pore volume increases[27].

*2.3 Scaling of morphological parameters*

The analysis of AFM topographic maps shows that the evolution of morphological parameters of ns-TiO$_2$ films is governed by simple scaling laws, known as the Family-Vicsek scaling relations, which describe a variety of growing interfaces[37–39].

Assuming that the vertical width of the interface is described by twice the rms roughness *w* and that the lateral dimension of the system is $L_0$ (Figure 2), the temporal and spatial scaling of *w* obeys the scaling relation:

$$w(L_0, t) \sim t^\beta f(L_0/\xi) \quad (1)$$

where the scaling function *f* has asymptotic behaviour:

$$f(x) = \begin{cases} x^\alpha, & x \ll 1 \\ 1, & x \gg 1 \end{cases} \quad (2)$$

According to Eqs (1) and (2), at a given time t, *w* increases as $L_0^\alpha$ then, above a crossover length $\xi$, saturates to a value proportional to $t^\beta$. The crossover length $\xi$, known as correlation length, scales as $t^{1/z}$. For a fixed lateral size $L_0$ of the system, w increases with time as $t^\beta$, then saturates to a value proportional to $L_0^\alpha$, after a crossover time that scales as $L_0^z$. The three exponents α, β, and z are called the roughness, growth, and dynamic exponents, respectively, and satisfy the relation $z = \alpha/\beta$. These exponents are peculiar of different growth models and mechanisms, so in principle, by determining two out of three of them, it is possible to get insights on the physical mechanisms governing the growth of the film[37,39]. More practically, the characterization of the scaling exponents allows predicting, and controlling, the evolution of the surface morphology of nanostructured films.

Typically, w is calculated from AFM topographic maps acquired on films with different thickness h (i.e. produced varying the deposition time t, at constant deposition rate, so that h ~ t), and β is obtained by a linear regression of the experimental curve $w \sim h^\beta$ on a loglog scale (Figure 5(a)). According to our data, the growth exponent of cluster-assembled ns-TiO$_2$ films is β = 0.38 ± 0.04, for Ar, and β = 0.39 ± 0.02 for He. The scaling of the surface rms roughness is therefore independent on the carrier gas (although the absolute value of w, at a given deposition time, can be dramatically different, as shown in Figures 4(a,b) and, quantitatively, in Figure 5(a)).

The experimental characterization of the roughness exponents α can be done in principle according to Eq (1) by applying a linear regression on a loglog scale to the w(l) experimental curve in the scaling region $l \ll \xi$ or, alternatively, to the height-height correlation curve $G(l) = \langle [h(x)-h(x+l)]^2 \rangle_x$, owing to the fact that:[37–39]

$$G(l) = \begin{cases} l^{2\alpha}, & l \ll \xi \\ 2w^2, & l \gg \xi \end{cases} \qquad (3)$$

This characterization however presents some difficulties, due to the fact that a) the width of the linear region (in a loglog scale) is limited to fractions of ξ, i.e. to few tens of nanometers in the case of ns-TiO$_2$, therefore extracting α by a linear regression in the scaling region could provide inaccurate results; b) only the asymptotic behavior of the scaling function G(l) is known, and existing analytical models of G(l) are only approximate in the crossover region l ≈ ξ (see for instance Refs[40–42]), therefore a nonlinear regression aimed at extracting *simultaneously* α and ξ could provide inaccurate results.

Our analysis strategy was therefore twofold. On one side, we *independently* measured ξ by applying a linear regression in semilog scale to the height correlation function C(l) = <h(x)h(x+l)>$_x$, for which we approximately assumed the scaling typical of a Gaussian surface[42]:

$$C(l) = w^2 e^{-l/\xi} \quad (4)$$

The advantage of Eq (4) is that in principle it allows extracting ξ as the inverse slope of a linear regression across a large spatial range, rather than as a free parameter of a multi-parameters approximated nonlinear model of G(l). Having independently measured w and ξ, we calculate the experimental reduced height-height correlation function G(l/ξ)/(2w$^2$), so that α can be obtained by a nonlinear regression of data using the model for G(l) proposed in Ref. [42]:

$$G(l) = 2w^2 \left[1 - e^{-(l/\xi)^{2\alpha}}\right] \quad (5)$$

with α as the only free parameter (Figure 5(c)). Eventually, once α and β are known, the dynamic exponent z can be calculated as z = α/β. Following this method we have obtained for the roughness and dynamic exponents of cluster-assembled ns-TiO$_2$ films the following results: α = 0.87 ± 0.04 and z = 2.29 ± 0.12, for Ar, and α = 0.86 ± 0.02 and z = 2.20 ± 0.12 for He. As in the case of the exponent β, the same scaling is observed irrespective to the carrier gas used, an evidence of the existence of general growth mechanisms governing the evolution of these interfaces.

On the other side, the exponent z (or better the quantity 1/z) can be directly characterized by a linear regression in loglog scale of the experimental curve ξ ~ h$^{1/z}$ (Figure 5(b)), which describes the evolution of the lateral width of the surface. Then, the roughness exponent α can be calculated as α = β/(1/z). The agreement of the values of α and z determined by the two independent methods is a test of the validity of the Family-Vicsek scaling relation z = α/β. Following this alternative method, we have obtained the following results: $1/z = 0.43 \pm 0.02$ ($z = 2.33 \pm 0.11$) and $\alpha = 0.88 \pm 0.06$ in the case of Ar; $1/z = 0.39 \pm 0.05$ ($z = 2.56 \pm 0.30$) and $\alpha = 1.00 \pm 0.14$ in the case of He. The values of the roughness and dynamic exponents obtained by the two independent methods agree within the experimental errors (in the case of He, the characterization of the scaling of the correlation length with thickness is clearly affected by the poor statistics of the sample, as witnessed by the large experimental error, and this uncertainty propagates to the value of α).

In summary, taking the weighted average of the values of α and z obtained by the two methods, we have obtained for the scaling exponents of ns-TiO$_2$ films the results reported in Table 1.[1]

Table 1. Scaling exponents. Average values of the scaling exponents of morphological parameters (Eqs. 1,2) for ns-TiO2 films deposited using Ar and He as carrier gas in the cluster source.

|   | Ar | He |
|---|---|---|
| α | 0.87 ± 0.03 | 0.86 ± 0.02 |
| β | 0.38 ± 0.04 | 0.39 ± 0.02 |
| z | 2.31 ± 0.08 | 2.56 ± 0.30 |

Values of the scaling exponents α and β similar to those of cluster-assembled ns-TiO$_2$ films are generally reported in literature in the context of percolative models dealing with liquid fronts advancing inside porous media, or with the propagation of burning fronts[37,39,44–47]. As previously noted for the scaling of nanostructured carbon assembled by SCBD[48,49], a common aspect is probably represented by the presence of pinning centers (quenched noise), which alter the advancement of the evolving front, in this case the evolution of the film surface during the growth process. The results of numerical simulations of the growth of a granular medium by deposition of size-distributed disks[50] are similar to ours for what concerns the roughness exponent α and to a lesser degree the growth exponent β. In all these cases the presence of a distribution of particle sizes represents the key factor, because the large particles in the tail of the distribution can act as a source of quenching noise. Concerning the value of the growth exponent β, it is compatible with a ballistic deposition model (for which β = 0.33), where incoming particles land on the growing interface, stick, and do not diffuse significantly[37,39].

---

[1] Sampling effects may cause inaccurate estimation of the scaling exponent α[43]; this effect has been studied and quantified for N = 512 points per line; here N = 2048, so minor effects can be expected, probably within the experimental error.

As the rms roughness w and the correlation length ξ, also the specific area r is an increasing function of film thickness, and its growth is governed by simple power laws. The scaling of the surface specific area excess $\Delta r = r-1$ with film thickness, $\Delta r \sim h^\delta$ is shown in Figure 5(d). We have found δ = 0.27 ± 0.01 for Ar, and δ = 0.45 ± 0.04 for He. In the case of He as carrier gas, the specific area excess grows faster; this behavior can be understood in terms of the smaller average particle size with respect to those produced in Ar, leading to a more compact interface, with smaller sub-nanometer pores and higher surface-to-volume ratio. Concerning the *absolute* value of specific area, in the accessible thickness range, for a given thickness, this is larger for films deposited using Ar (which is clear also from Figures 4(a,b)). Specific area is closely related to the local surface slope, which is expected to follow a similar trend (while the *mesoscopic* average slope $s_{meso} \approx 2w/\xi$ and specific area $r_{meso} \approx 1+2(w/\xi)^2$ parameters, evaluated on a scale ξ, are not expected to change much, because w and ξ scaling exponents are similar).

*2.4 The impact of morphological parameters on relevant interfacial phenomena*

The great advantage of SCBD for the fabrication of nanostructured materials by cluster-assembling is that it allows controlling precisely the morphological properties during the growth process, by simply varying the deposition time. The hard work is done by the ballistic deposition process and the underlying growth mechanisms, whose practical consequences are well described by the scaling laws discussed in the previous sections. Cluster-assembled nanostructured substrates deposited by SCBD can therefore be used as platforms for the study and the control of several interfacial properties like wettability, charging in liquid electrolytes and formation of electrostatic double layers, as well as of relevant phenomena like protein adsorption, and cell adhesion and proliferation. On the basis of the information acquired in this study on the scaling of the surface morphology of nanostructured titania films and of the results of previous studies, we here discuss the impact of nanoscale morphological parameters on the formation of an electrostatic double layer at the interface between ns-TiO$_2$ and a liquid electrolyte, and the relevance of surface morphology and charging on the adsorption of proteins and nanoparticles.

*2.5 Surface morphology and electrostatic double layers*

Surface morphology and nanopores play an important role in processes involving the interaction of entities (protein, viruses, nanoparticles) with ns-TiO$_2$ and other nanostructured surfaces, via the modulation of electric interfacial properties. In particular, when the nanostructured material is used to produce electrodes and substrates for operation in liquid electrolytes, with given pH and ionic strength, electrostatic double layer phenomena take place,[51–53] causing the onset of an electric charge distribution at the solid-liquid interface and of an electrostatic potential that vanishes into the bulk of the electrolyte. Double-layer phenomena have been recently shown to be strongly influenced by the morphological properties of the surface[54–58]. Thanks to the control of morphological parameters provided by SCBD, we have recently performed a systematic exploration of the electrical double layer properties in different interaction regimes, characterized by different ratios of characteristic lengths of the system: the Debye screening length $\lambda_D$ of the electrolyte, the surface rms roughness w and the correlation length ξ and, on a local scale, pores' width and depth.

The value of pH at which the net surface charge is zero, the isoelectric point (IEP), plays an important role in the interaction of the surface with particles like proteins, enzymes or catalysts[59]. Upon observation of a remarkable reduction by several pH units of the IEP on rough ns-TiO$_2$ nanostructured surfaces with respect to flat crystalline rutile TiO$_2$, we proposed that roughness-induced self-overlap of the electrical double layers can be responsible for deviations from the trend expected for flat surfaces[58]. Nanostructured materials offer the unique possibility of inducing self-overlap effects between adjacent regions of the same surface, due to the abundance of high aspect-ratio nanoscale regions providing extreme confinement to the electrolyte. At the liquid-solid interface, due to the very large curvature of the surface and the nanoporous nature of it, the self-overlap effect can be dramatic, leading to severe charge-regulation phenomena and strong deviations from the linearized Poisson-Boltzmann theory of the electrostatic double layer[52].

Simple geometrical arguments suggest that the relative degree of self-overlap γ of electrostatic double layers within a pore, defined as the ratio of the total volume of the double layer (up to a distance $\lambda_D$) to the volume of the overlapping region, is an increasing function of the ratios $\lambda_D/\xi'$ and $s = w'/\xi'$, where $\xi'$ and w' are pore's half-width and depth, respectively, so that s is the pore's slope[58]:

$$\gamma = \begin{cases} \frac{(\lambda_D/\xi')s}{2\sqrt{1+s}-(\lambda_D/\xi')s} & , s \leq 1 \\ \frac{(\lambda_D/\xi')s}{2\sqrt{1+s^2}-(\lambda_D/\xi')s} \frac{1}{2}\left(1+\frac{1}{s^2}\right) & , s > 1 \end{cases} \quad (6)$$

Here we have calculated the factor γ for two representative populations of surface pores of two ns-TiO$_2$ films with thickness of 50nm and 340nm (low- and high-roughness limits, respectively). According to previous studies[27], the distribution of aspect ratios of single surface pores turned out to be shifted towards higher values in thicker samples. Thick films also showed a higher number of pores with smaller width, and the pores with the highest aspect ratio turned out to be preferentially those with smaller width. Both elements determine an increase of γ according to Eq. 6. This can be appreciated in Figure 6 where the distributions of γ values of the two films are compared. The median value of γ increases from 5.9% to 7.2% as the film thickness increases. Moreover and most noticeable, the fraction of pores with values of gamma in the interval 20-60% increases of about 40% as a consequence of surface roughening. What are not represented in Figure 6 are the populations of smaller pores, with width comparable to the AFM tip radius (i.e. smaller than approximately 10nm), because of the intrinsic resolution limits of the microscope. Likely, the self-overlap of the double layer in these pores is almost complete, because the pores' dimensions are smaller than the Debye length. Also in the case of more concentrated electrolytes, like PBS (ionic strength about 150-200 Mm) with $\lambda_D$<1nm, there will be a significant fraction of surface pores with size comparable to $\lambda_D$ where γ≈1. We therefore expect that the real distributions of γ values are significantly shifted towards higher values with respect to those shown in Figure 6, and that the difference between the distributions of thin and thick ns-TiO$_2$ films can be significantly larger than that shown in Figure 6.

*2.6 Influence of surface morphology and charging on the adsorption of proteins and nanoparticles*

Nanostructured surfaces promote the formation of protein aggregates, because nanometric pores generate the conditions for protein nucleation inside the pores, where they get trapped and mutually interact. This mechanism accounts for the observed improved adsorption of several different proteins, including enzymes, on ns-TiO$_2$ films, well beyond expectations due to the geometrical enhancement provided by the increased specific area[27,28]. Moreover, it has been observed during experiments on the adsorption of the enzyme trypsin on ns-TiO$_2$ that the aggregation of the protein inside nanopores perturbs its relative catalytic activity in a roughness-dependent manner. In particular, the specific activity per mass of adsorbed enzyme decreased linearly with roughness, reflecting the reciprocal increase of steric hindrance of active sites with the ns-TiO$_x$ roughness[28]. These observations are of relevance for a broad class of surface-active nanoparticles, from biological enzymes to solid-state nano-catalysts.

Based on the reported experimental evidence and considerations, we present in Figure 7 a tentative simplified scheme of the protein or nanoparticle adsorption process on ns-TiO$_2$ and other nanostructured surfaces, taking into account the direct effect of nanoscale morphology, via nucleation into nanopores[27], as well as the indirect effect, via electrostatic double-layer interactions (and dispersion interactions). At large distance (d >> w,$\lambda_D$, box A) nanoparticles diffuse in the liquid buffer almost freely. They can feel weak electrostatic interactions depending on their net charge and on the net charge of the surface (i.e. on their IEP, on the IEP of the surface, and on the pH of the buffer). The surface charge of the surface can be thought to be uniformly distributed on an average plane of charge[58], lying at a distance equal to the rms roughness w below the most protruding asperities (representing the points of first contact for incoming species). At distances comparable to the extension of the double layer (d ≈ $\lambda_D$, box B), electrostatic interactions become stronger and may attract or repel particles, depending on the net sign of charges (metal oxide surfaces are typically negatively charged in water). In the case of an effective, attractive net interaction, particles are driven towards the surface. When the condition d ≈ w is also satisfied, particles may feel inhomogeneous electric field, because the local curvature of the surface can influence the surface charge distribution and surface potentials. Strongest effect will be observed when the double layer resides all within the surface, i.e. when d ≈ w and w ≈ $\lambda_D$. Particles at this stage tend to migrate towards the surface asperities, where the electric field is more intense; van der Waals interactions at this stage are still weak and do not contribute significantly[60]. When particles approach at a distance smaller than the rms roughness (d < w, box C), dispersive interactions become important and drive them deeply into the pores, where conditions are met to start nucleation. The particle–surface interaction is now influenced by steric, orientational effects depending on shape and size of particles and surface features, whose finer granularity comes into play. The average plane of charge description no longer holds; the distribution of charged residues on the particle's surface, as well as the local variations of surface charges, potentials, ionic concentration on the surface, drive the interaction. From this moment on (d << w, box d), the interaction regime definitely changes from particle/surface to particle/particle, which definitely boosts nucleation inside pores and determines the increased particle loading capacity of the nanostructured surface compared to smoother surfaces (actually not necessarily flat, but lacking the nanoscale structuring, and the multi-scale hierarchical organization of surface features).

3. **Conclusions**

SCBD coupled to a Pulsed Micro-plasma Cluster Source is an additive technology allowing the deposition of nanostructured titanium dioxide films, with finely-tuned morphological properties. We have shown here that the very good control over film surface nanoscale morphology is achieved by taking advantage of simple scaling laws governing the ballistic deposition regime of low-energy, mass-dispersed clusters with reduced surface mobility. We have analysed several hundred different samples produced over a time interval of ten years with different deposition apparatus based on SCBD. We quantitatively demonstrated that SCBD is a robust approach for the large-scale production of nanostructured surfaces with reproducible nanoscale morphology.

The possibility of controlling the complex surface morphology of titania thin films allowed investigating diverse morphology-dependent interfacial properties, such as wettability, electrostatic double-layer and surface charge density, and protein and enzyme adsorption. Functional properties of nanostructured surfaces turn out to be finely regulated by the competition of different characteristic nanoscale lengths of the system, which in turn can be tuned and prototyped by exploiting SCBD. These characteristic lengths also match the typical dimensions of the objects interacting with the surface, either biomolecules (proteins, enzymes) or nanoparticles (catalysts, quantum dots, plasmonic particles, etc.). It is clear therefore that for applications in nanobiotechnology SCBD is an enabling technology for the high-throughput fabrication and engineering of surfaces.

**Acknowledgements**

We thank P. Piseri, G. Bongiorno, E. Sogne, Fondazione Filarete, Tethis spa for support in the deposition of ns-TiO$_2$ films. We thank V. Vyas, V. Cassina, G. Berlanda and M. Galluzzi for support in AFM characterization.

**Figure captions**

**Figure 1.** The SCBD apparatus. The mixture of gas and clusters is accelerated by a difference of pressure between the interior of the cluster source (higher pressure) and the expansion chamber (lower pressure, high vacuum), and collimated through the aerodynamic focuser. A skimmer selects the central portion of the beam. Eventually, nanoparticles enter the deposition chamber and are deposited on a substrate to form a film with thickness in the 1-1000 nm range. The details of the PMCS are shown in the inset: the pulsed valve (1) injects the carrier gas into the ceramic body (2) of the source, hosting the Ti rod (3); the focuser (4) containing the aerodynamic lenses concentrates the nanoparticles along the beam axis.

**Figure 2.** Schematic representation of a topographic one-dimensional profile h(x) and of relevant morphological quantities: rms surface roughness $w$, length L and projected length $L_0$, the (halved) average lateral dimension $\xi$ of surface features (correlation length), and the mean height $h_0$.

**Figure 3.** Representative AFM topographic maps (top views) of ns-TiO$_2$ clusters deposited on smooth amorphous silica substrates (glass coverslips) using He (a,b) and Ar (c,d). Clusters from the center (a,c) and from the periphery (b,d) of the beam have been selected for deposition; in boxes (e-h) the corresponding distributions of particles' diameters in semilog scale are shown.

**Figure 4.** Representative topographic maps of ns-TiO$_2$ films. (a,b) Similar thickness (about 100 nm) but different carrier gas (He and Ar, respectively); the value of rms roughness is 8 and 21 nm, respectively. (c,d) Same carrier gas (Ar) but different thickness (the value of h is about 20 and 100 nm, respectively); rms roughness w is 10 and 22 nm, respectively. Representative surface profiles are superimposed to 3D views in (c,d) to highlight nanometric pores of diverse depths and widths. The vertical scale range for all images is 100 nm.

**Figure 5.** Scaling of morphological properties of ns-TiO$_2$ films deposited using He and Ar as carrier gas. (a) rms roughness $w \sim h^{\beta}$; (b) the lateral width of the surface $\xi \sim h^{1/z}$; (c) the reduced height-

height correlation function $G(\Delta x/\xi)/(2w^2)$; (d) the surface specific area excess $\Delta r = r-1 \sim h^\delta$. All curves are plotted in loglog scale. Linear regressions in loglog scale have been applied to extract the scaling exponents from curves in (a,b), and (d), while Eq. 5 has been fitted to data in (c).

**Figure 6.** The self-overlap coefficient γ for electrostatic double layers between adjacent regions of the same surface, calculated from experimental morphological data for ns-TiO$_2$ films with thickness of 50 and 340 nm (SMP1 and SMP5 in Ref. [27]).

**Figure 7.** A tentative simplified scheme of the protein adsorption process on nanostructured surfaces, taking into account the direct effect of nanoscale morphology, via nucleation into nanopores, as well as the indirect effect, via electrostatic double-layer (and dispersion) interactions.

**Figures**

Figure 1

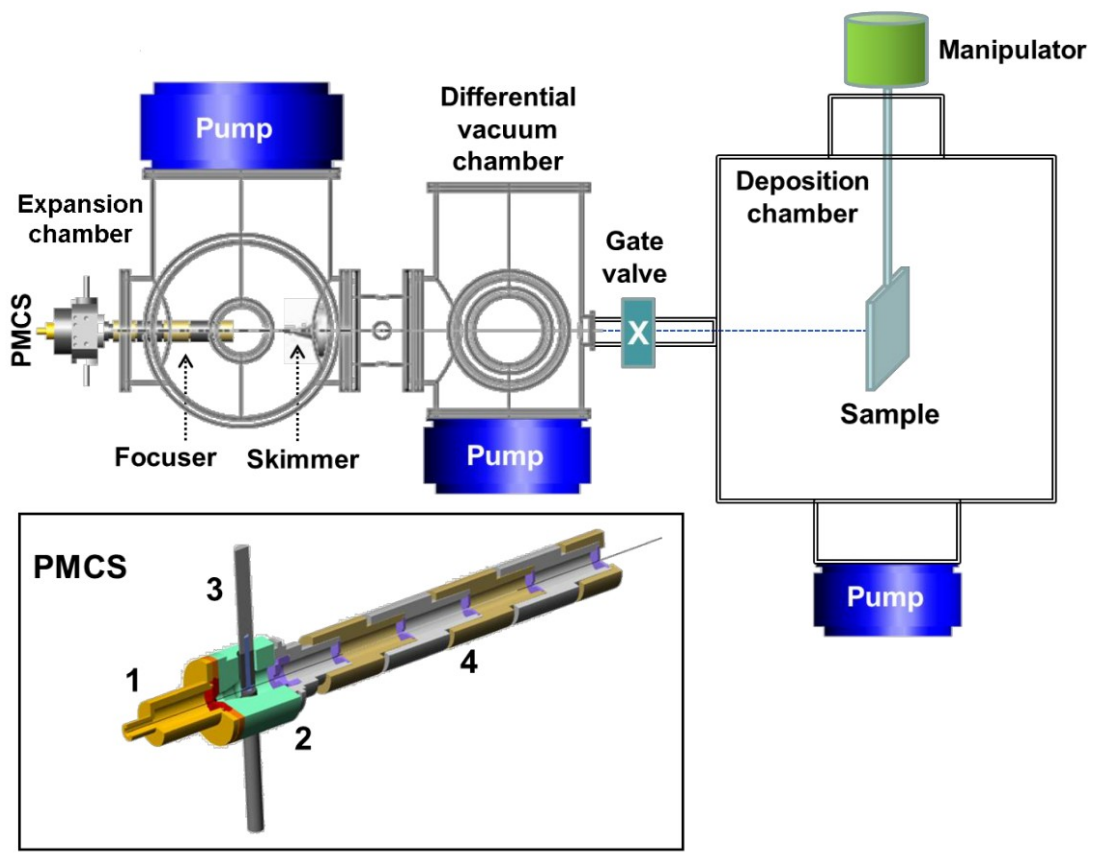

Figure 2

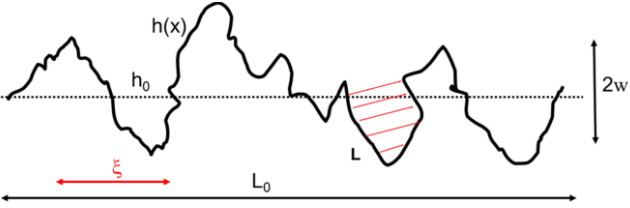

Figure 3

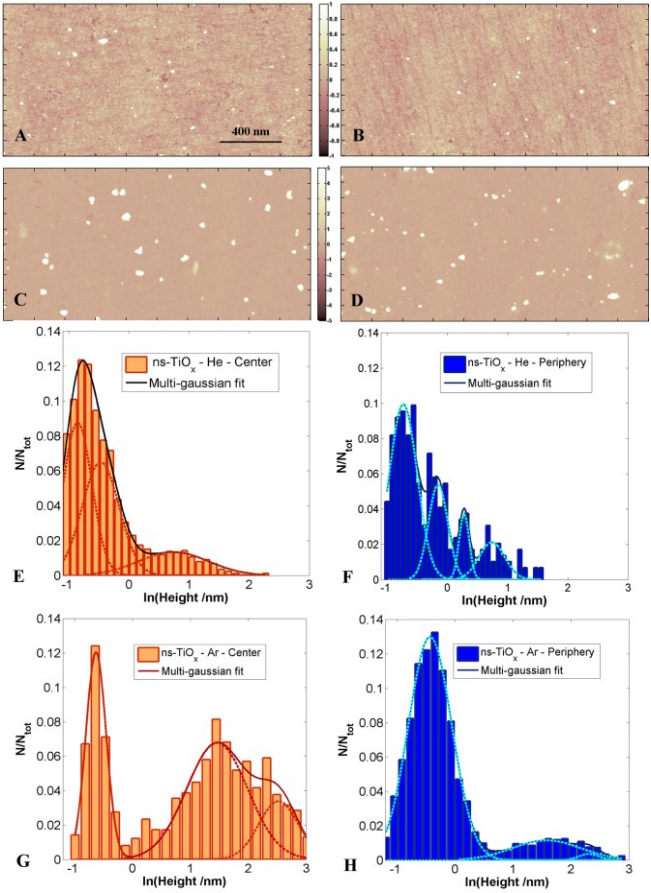

Figure 4

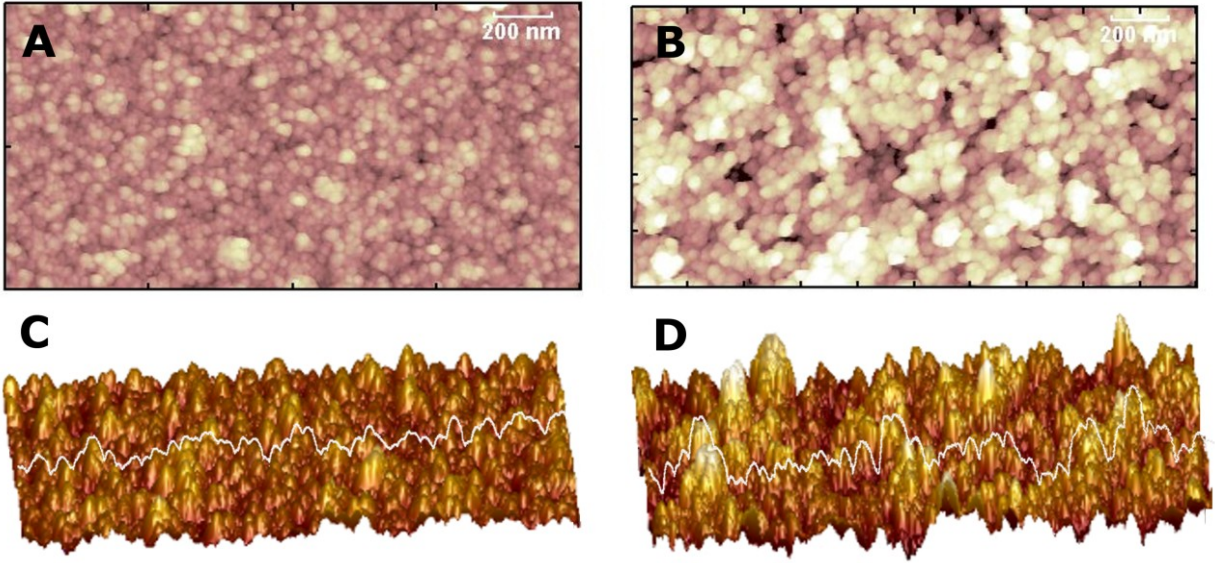

Figure 5

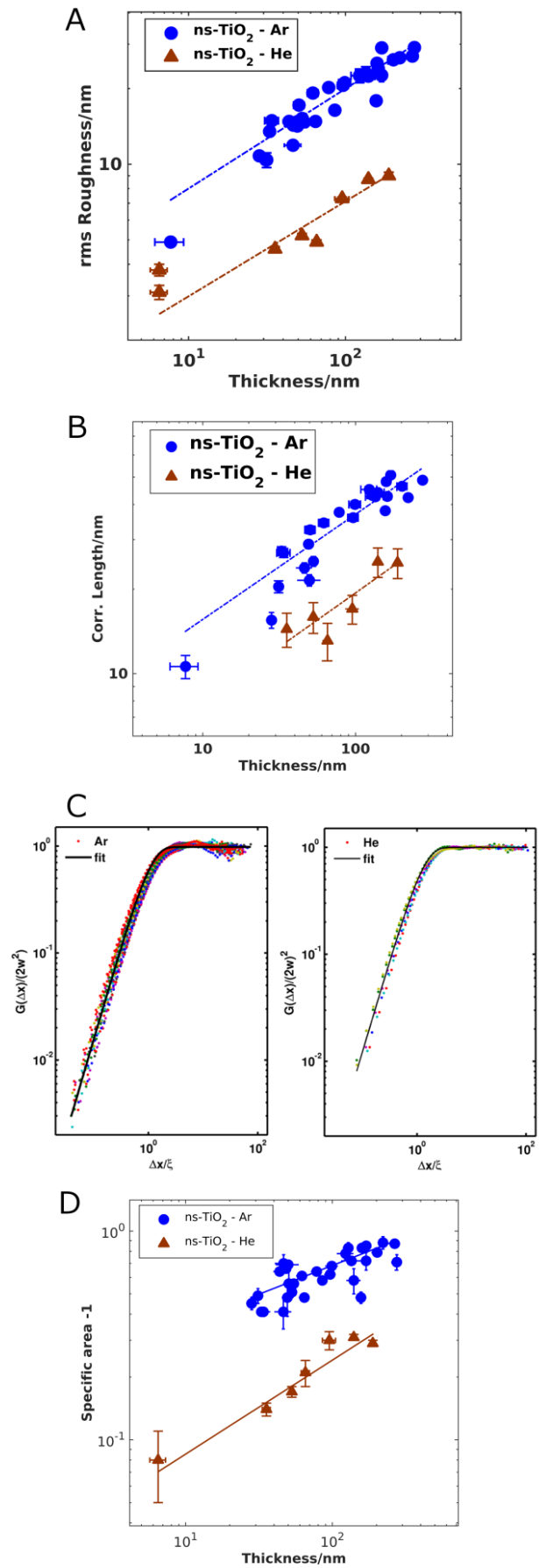

Figure 6

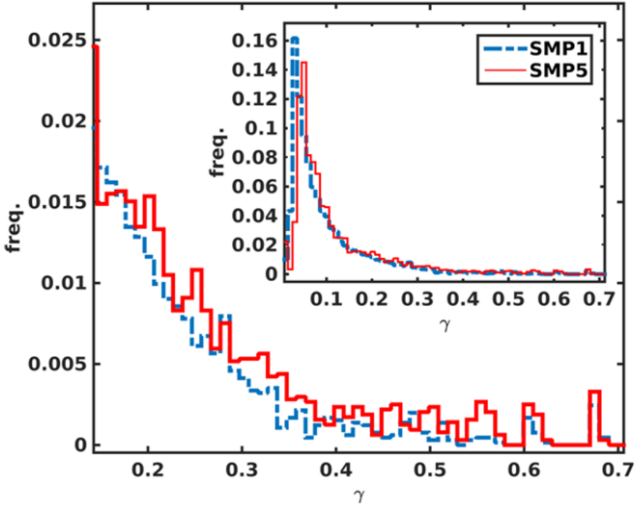

Figure 7

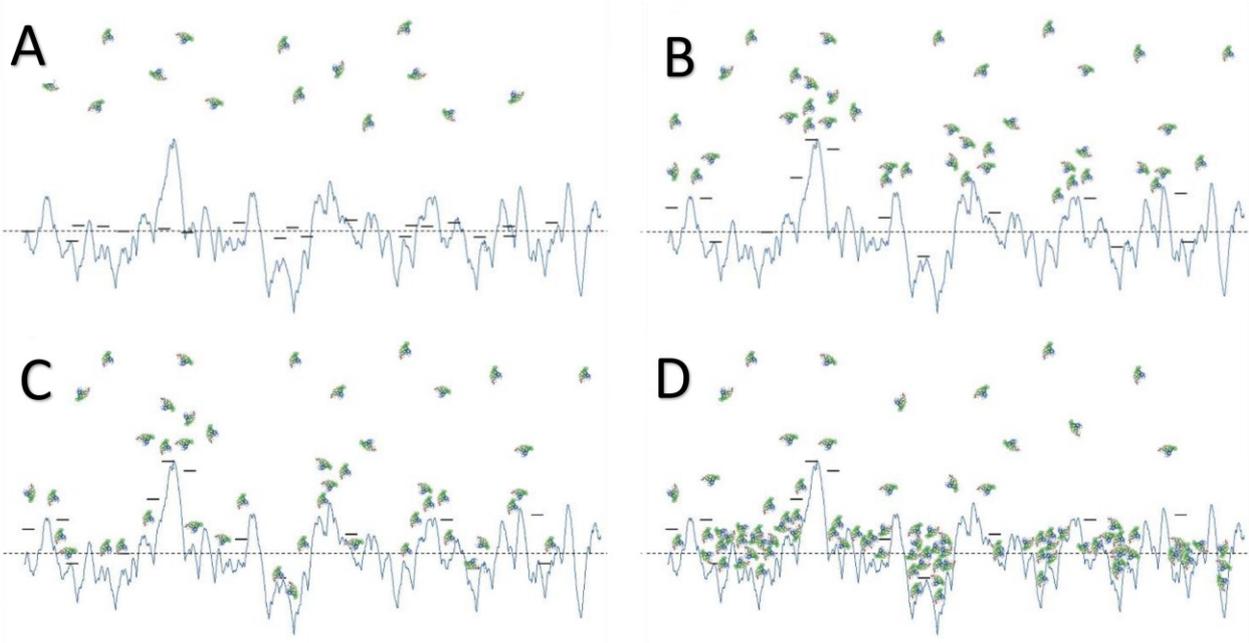